# Identifying outstanding transition-metal-alloy heterogeneous catalysts for the oxygen reduction and evolution reactions via subgroup discovery

Lucas Foppa* and Luca M. Ghiringhelli

The NOMAD Laboratory, Fritz-Haber-Institut der Max-Planck-Gesellschaft, Faradayweg 4-6, D-14195 Berlin, Germany; Humboldt-Universität zu Berlin, Zum Großen Windkanal 6, D-12489 Berlin, Germany.

**Abstract:** In order to estimate the reactivity of a large number of potentially complex heterogeneous catalysts while searching for novel and more efficient materials, physical as well as data-centric models have been developed for a faster evaluation of adsorption energies compared to first-principles calculations. However, *global* models designed to describe as many materials as possible might overlook the very few compounds that have the appropriate adsorption properties to be suitable for a given catalytic process. Here, the subgroup-discovery (SGD) *local* artificial-intelligence approach is used to identify the key descriptive parameters and constrains on their values, the so-called SG rules, which particularly describe transition-metal surfaces with outstanding adsorption properties for the oxygen reduction and evolution reactions. We start from a data set of 95 oxygen adsorption energy values evaluated by density-functional-theory calculations for several monometallic surfaces along with 16 atomic, bulk and surface properties as candidate descriptive parameters. From this data set, SGD identifies constraints on the most relevant parameters describing materials and adsorption sites that (i) result in O adsorption energies within the Sabatier-optimal range required for the oxygen reduction reaction and (ii) present the largest deviations from the linear scaling relations between O and OH adsorption energies, which limit the performance in the oxygen evolution reaction. The SG rules not only reflect the local underlying physicochemical phenomena that result in the desired adsorption properties but also guide the challenging design of alloy catalysts.

**Keywords:** artificial intelligence, subgroup discovery, symbolic inference, supervised descriptive rule induction, transition-metal surfaces

## Introduction

Among the multiple processes that govern heterogeneous catalysts,[1-3] the bond-breaking and -forming reactions occurring on the catalyst surface, and, in particular, the associated (free-) energy barriers, play an important role in determining the reactivity of a given material. The energy barriers of surface reactions have been related to the adsorption energy of reactants, reaction intermediates or products via linear Brønsted-Evans-Polanyi relationships.[4,5] Adsorption energies can be evaluated using *ab initio* methods, for instance via density-functional-theory (DFT) calculations. However, the explicit evaluation of adsorption energies by accurate first-principles methods for a large number of materials, desirable in the context of catalyst screening approaches, becomes impractical when complex catalysts such as transition-metal alloys are considered. This is because these materials display a large number of possible surface sites which could play a role in catalysis.

In order to efficiently explore a large number of possibly complex materials in the quest for novel catalysts, the scaling relation approach,[6] among other physical[7] or data-centric[8] models, have been used for the estimation of adsorption energies at lower computational effort compared to DFT. The scaling relations exploit the approximately linear relationships between adsorption energies of different surface species to reduce the number of explicit DFT calculations needed to investigate a certain catalytic process. Such linear models are designed to estimate adsorption energies for as many different materials and surface sites as possible. However, only very few of the investigated systems present the appropriate adsorption properties to be useful for a given catalytic process. Firstly, the adsorption energies of key reaction intermediates typically need to lie in a Sabatier-optimal range for the performance to be maximized.[9-11] Secondly, the adsorption energies of different species might need to be tuned independently for an optimal reactivity to be achieved.[12] This implies that deviations from the linear relationships between adsorption energies, which describe the trend for most of the materials, might be actually desirable.[13] In both these situations, the interesting materials and surface sites thus present *statistically exceptional* adsorption properties. This questions the suitability of using global models to screen for new catalysts.

Here, we apply the subgroup-discovery (SGD) artificial-intelligence *local* approach[14-19] to identify key descriptive parameters - and constraints on their values-, which are particularly associated to outstanding adsorption properties of transition-metal surfaces. In particular, we introduce a strategy to address target properties whose desired values lie in a specific range and use this approach to describe adsorption sites presenting Sabatier-optimal oxygen adsorption energies for the oxygen reduction reaction (ORR).[20] Additionally, we show how SGD can be used to describe data points that deviate the most from a given model such as the linear scaling relations between O and OH adsorption energies on different surface sites. Such scaling relations impose a limit for the optimization of oxygen evolution reaction (OER) performance.[21] Thus, materials and adsorption sites deviating from the linear scaling are the interesting ones. The ORR and the OER are two crucial processes for energy conversion and storage.

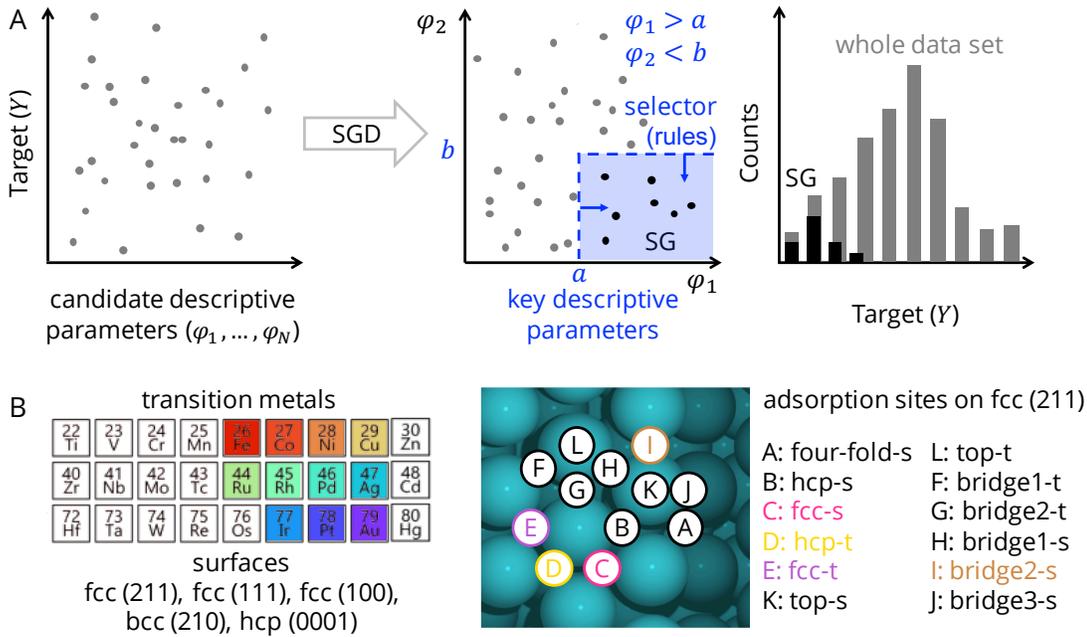

**Figure 1.** A: Illustration of the SGD approach for identifying key descriptive parameters and rules determining SGs with outstanding distribution of the target. The rules are constraints on the values of key descriptive parameters. The distribution of target values in the SG might be outstanding because it is, for instance, narrower than the distribution of the target values over the whole data set. B: Transition metals and surfaces considered in this work. We consider the face-center-cubic (fcc) structure for all metals except Fe, for which the body-center-cubic structure (bcc) and the (210) surface is considered. For Co, the (0001) surface of the hexagonal-closed-packed structure is also included. The adsorption sites for the fcc (211) surface are also shown in detail on the right. This surface termination contains both terrace- and step-edge-like sites, labelled "t" and "s" in the figure, respectively.

**Subgroup discovery approach**

We start our analysis by introducing the SGD approach[14-18] to uncover complex patterns associated to outstanding local behavior by using data sets. This methodology has been recently applied to catalysis[22] as well as materials-science[18,23] problems.

The SGD method is based on an input data set, which we refer to as the *population P* of data points, each of them associated to a different material or, in the case of this work, to a different surface site. For each of the data points, the value of a *target* of interest, $Y$, and the values of $N$ potentially-relevant *candidate descriptive parameters*, denoted $\varphi_1, \varphi_2, \ldots, \varphi_N$, are known. The candidate descriptive parameters are structural or physicochemical parameters that possibly correlate with the target. Starting from such data set, SGD identifies subsets of data, hereafter *subgroups* (SGs), that present an outstanding distribution of the target values with respect to the whole data set (Fig. 1A). The so-called *quality function* $Q(P, SG)$ measures how outstanding a SG is compared to the whole data set. This function typically has the form

$$Q(P, SG) = \frac{s(SG)}{s(P)} * u(P, SG), \quad (1)$$

where the first term, the *coverage*, contains the ratio between the number of data points $s$ in the subgroup and the total number of data points in the whole data set. The coverage controls the subgroup size and prevents that very small SGs with little statistical significance are selected. The second term $u(P, SG)$, the *utility function*, measures the dissimilarity between the SG and the population. It can be chosen[18] depending on the scientific question of interest (vide *infra*).

The SGD algorithm consists in two steps. Firstly, combination of statements (hereafter *selectors*, $\sigma(\varphi)$) about the data are generated. The selectors are Boolean functions defined through conjunctions of propositions and have the form

$$\sigma(\varphi) \equiv \pi_1(\varphi) \wedge \pi_2(\varphi) \wedge \ldots \wedge \pi_p(\varphi), \quad (2)$$

where "∧" denotes the "and" operator and each proposition $\pi_i$ is, for instance, an inequality constraint on one of the descriptive parameters

$$\pi_i(\varphi) \equiv \varphi_i \geq v_j \text{ or } \pi_i(\varphi) \equiv \varphi_j < v_j, \quad (3)$$

for some constant $v_j$ to be determined during the analysis. The *selectors* describe convex regions in the descriptive parameter space defining the SGs. To keep the number of $v_j$ values computationally tractable, a finite set of cut-offs is determined using k-means clustering, where k is a parameter to be assessed. Secondly, a Monte Carlo search algorithm is used to find SGs, defined by the selectors generated in the first step, that maximize the quality function. The most relevant SGs are those for which the quality function reaches the highest values. The selectors defining those SGs, and, more specifically, the propositions in the selectors, contain the key descriptive parameters associated to the underlying processes that exclusively govern the local behavior within the subsets (or SGs) of data points. The propositions entering the selectors can be thus seen as *rules* determining the outstanding SG behavior. Therefore, the SG is at the same time the subset of selected data and the selector, i.e., the rules that are used to obtain this selection. In fact, the SG rules are more relevant than the particular subset of selected (training) data. Further SGD details are available in Electronic Supporting Information, ESI.

## Data set of adsorption energies and candidate descriptive parameters

We analyze a dataset containing 95 oxygen (atomic O) adsorption energies, which were calculated with DFT using the van der Waals-corrected BEEF-vdW exchange-correlation functional in previous contributions.[8,24] Eleven transition metals and several adsorption sites of different surfaces for which (meta)stable oxygen adsorption is observed were included in our analysis (Fig. 1B). The oxygen adsorption energy is defined as

$$E_{ads}^{O} = E_{surf,clean} + 0.5\, E_{O_2(g)} - E_{surf,ads}, \quad (4)$$

where $E_{O_2(g)}$, $E_{surf,clean}$ and $E_{surf,ads}$ are the total energies of the O$_2$ gas-phase molecule, clean surface, and surface containing the O adsorbate, respectively. Positive oxygen adsorption energy values correspond, therefore, to favorable adsorption with respect to the gas-phase molecule.

**Table 1.** Candidate descriptive parameters used for the SGD of outstanding transition-metal catalysts.

| type | | description | Ref. |
|---|---|---|---|
| atomic | PE | Pauling electronegativity | [25] |
| | IP | ionization potential | [26] |
| | EA | electron affinity | [26] |
| bulk | bulk$_{nnd}$ | nearest-neighbor distance | [8][a] |
| | $r_d$ | $d$-orbital radius | [27] |
| | $V_{ad}^2$ | coupling matrix element between the adsorbate states and the metal $d$ states squared | [28] |
| surface | $W$ | work function | [8][a] |
| surface site | site$_{no}$ | number of atoms in the ensemble | [8][a] |
| | CN | coordination number | [8][a] |
| | site$_{nnd}$ | nearest-neighbor distance | [8][a] |
| | $\epsilon_d$ | $d$-band center | [8][a] |
| | $W_d$ | $d$-band width | [8][a] |
| | $f_d$ | $d$-band filling | [8][a] |
| | $f_{sp}$ | $sp$-band filling | [8][a] |
| | DOS$_d$ | density of $d$-states at Fermi level | [8][a] |
| | DOS$_{sp}$ | density of $sp$-states at Fermi level | [8][a] |

[a] as determined by DFT-BEEF-vdW.

An important aspect in SGD is the choice of candidate descriptive parameters. Following reference [8], we use, as candidate descriptive parameters, the atomic, bulk, and clean surface properties shown in Table 1. The atomic parameters are properties that only depend on the element. The bulk, surface and site parameters are related to the geometry and the electronic structure of either bulk metals, or the surfaces and their adsorption sites. The surface- and surface-site-related descriptive parameters were evaluated on (relaxed) clean surfaces, i.e., without the presence of the adsorbed species, in reference [8]. The surface-site parameters were calculated as averages over the metal atoms that compose the site ensemble. In total, 16 parameters uniquely describing each material and surface site are used. We note that the candidate descriptive parameter set includes properties proposed to describe overall trends in adsorption energies such as as the $d$-band center ($\epsilon_d$)[7] or coordination numbers (CN)[29] as well as many other, potentially relevant, parameters.

## Subgroups of surface sites with optimal range of oxygen adsorption energies for the ORR

To illustrate how SGD identifies the relevant descriptive parameters and the rules describing surface sites that bind a certain reaction intermediate with a specific range of binding strength, we start our analysis by identifying SGs of surface sites providing oxygen adsorption energy close to $E_{ads,opt}^{O} = 1.8$ eV. Based on DFT-derived potential energy surfaces describing the main proposed mechanisms of the ORR, adsorbed oxygen was identified as a key intermediate in this reaction and the oxygen adsorption energy value of 1.8 eV was related to the highest activity over a series of transition-metal low-index (111) surfaces.[11] To take into account that a *range* of oxygen adsorption energies around 1.8 eV might result in catalysts that maximize the performance, we define, for our SGD analysis, a target that assumes small values in a given window around $E_{ads,opt}^{O}$ and rapidly increases outside such interval. Among several possible choices of functions that would reproduce this behavior, we use a quadratic expression and consider [1.3,2.3 eV] window of optimal adsorption energy values. Our SGD target is thus defined by

$$\Delta^O = \left(\frac{E^O_{ads} - E^O_{ads,opt}}{0.5\ eV}\right)^2, \quad (5)$$

where $E^O_{ads}$ is the oxygen adsorption energy for an arbitrary surface site. The distribution of $\Delta^O$ over the data set is shown in Fig. 2A and 2B. We are interested in SGs of data points for which $\Delta^O$ assumes low values. As utility function, we use

$$u(P, SG) = \frac{std(SG)}{std(P)}, \quad (6)$$

where $std(SG)$ and $std(P)$ are the standard deviation of the distributions of the target in the SG and in the whole data set, respectively. By using the ratio of standard deviations in the utility function, we favor the selection of SGs that present narrow distribution of values for the target.

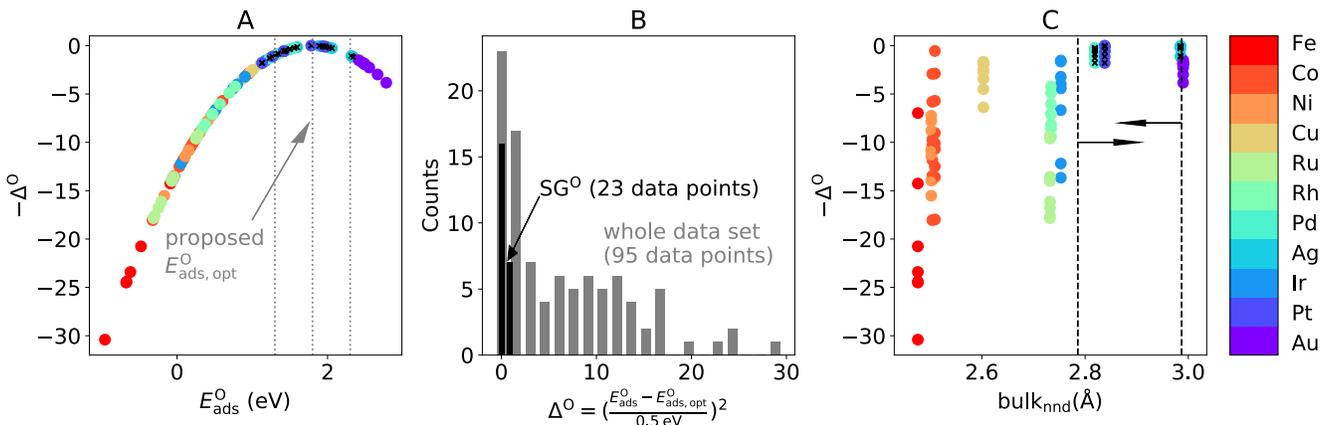

**Figure 2.** SGD of transition-metal catalysts presenting surface sites with an optimal range of oxygen adsorption energies. A: Visualization of the target quantity ($\Delta^O$), defined in Eq. 5, for the training data. $\Delta^O$, which is unitless, is smaller than 1 in an interval of $\pm 0.5$ eV centered around the proposed optimal value of $E_{ads,opt}^{O} = 1.8$ eV. B: Distribution of $\Delta^O$ in the whole data set and in the identified SG. C: SG selector, indicated by the dashed lines and by the arrows, on a identified key descriptive parameter: bulk nearest-neighbor distance (bulk$_{nnd}$). The data points corresponding to the SG are marked with black crosses in A and C.

Among the SGs that maximize the quality-function values, we identify a SG containing 23 data points, i.e., ca. 24% of the data set, which is relatively narrow and presents low target values (Fig. 2B, in black). Indeed, this SG contains the surface sites for which the oxygen adsorption energies are the closest to the proposed optimal value (Fig. 2A, in which the adsorption sites belonging to the SG are shown as black crosses). All considered adsorption sites of Pd, Ag and Pt surfaces are part of this SG. Pd and Pt are indeed known to be the best ORR catalysts among all metals included.[20] This SG is defined by the selector

$$\sigma^O = 2.786 < \text{bulk}_{nnd} \leq 2.987\ \text{Å}, \quad (7)$$

as shown in Fig. 2C. Therefore, the interatomic nearest-neighbours distance of the bulk materials is a key parameter determining if an adsorption site is associated to the optimal range of oxygen adsorption values for the ORR. In particular, $\text{bulk}_{nnd}$ needs to assume an intermediate range of values, given by (7), in order for a material to present surface sites with the desired oxygen binding strenght.

The SG rule given by (7) is the simplest SG rule identified, which only depends on one descriptive parameter. Several different SG rules (Table S1) result, however, in the exact same subselection of (training) data points and thus in the same quality-function values compared to the SG defined by (7). For instance, the selector

$$\text{site}_{nnd} > 2.759\ \text{Å} \wedge PE \leq 2.125, \quad (8)$$

which depends on two descriptive parameters, also selects the adsorption sites of Pd, Ag and Pt. The presence of similar SGs defined by slightly different rules is due to the fact that different descriptive parameters encode similar physicochemical information. Indeed, some of the candidate descriptive parameters are correlated with each other. In particular, the Pearson correlation between $\text{bulk}_{nnd}$ and $\text{site}_{nnd}$ is equal to 0.99 and between $\text{bulk}_{nnd}$ and PE is 0.72 (Fig. S3). We note that correlations involving more than two descriptive parameters, which are not captured by the Pearson correlation scores, might be also present within the training data set. This is not a limitation for SGD, since it can identify different equivalent descriptive rules (with respect to a given input training data).

By using the function defined in Eq. 5, we allow for SGD to focus on a range of desired target values based on the distance to the desired range. Another strategy to achieve this goal is to define a categorical target which labels the data points falling in the desired range of values. In this case, the data points inside or outside the desired range are treated equivalently irrespective of their distance to the optimal value or to the borders delimiting the range. To illustrate this approach, we have applied SGD using a categorical target indicating whether a surface site is at the optimal [1.3,2.3 eV] oxygen adsorption energy range (see details in ESI). Thus, the SGD is used to identify rules that *classify* the adsorption sites and materials as belonging to the Sabatier-optimal range. By using this approach, SGD identified similar rules compared to those derived using the numerical target $\Delta^O$. Additionally, the same subselection of data points as for the SG discussed in Fig. 2 is obtained. This is also the case when the optimal ranges of [1.4,2.2 eV] or [1.2,2.4 eV] are used, instead of [1.3,2.3 eV], for the labelling of data points. These results indicate that the SG rules are stable with respect to the choice of target function and optimal range of values around the proposed optimum (within the ranges ±0.4-0.6 eV).

We then exploited the rules defining the SGs of outstanding $\Delta^O$ values (all the rules shown in Table S1 for the $\Delta^O$ target), which were obtained using only monometallic surfaces, to select surface sites of transition-metal alloys that bind oxygen with the appropriate range of adsorption energies. Thus, we apply the constraint that the alloy surface sites of interest should simultaneously satisfy several of the SG rules identified using the monometallic systems. The evaluation of adsorption energies on surfaces of metal alloys is a particularly resource-consuming task for DFT, as the number of possible metal combinations and surface sites grows significantly with respect to the monometallic systems. We focus on bimetallic alloys with 1:1 atomic ratio, since the candidate descriptive parameters for 25 of such alloy compositions, i.e., the properties listed in Table 1, are available in reference [8]. In total, 360 different surface sites of (211) surfaces are considered. The alloy *atomic* descriptive parameters are taken as the average between the atomic properties of the two metals in their composition. The surface- and site-related parameters are explicitly evaluated by DFT using an alloy atomistic models in reference [8].

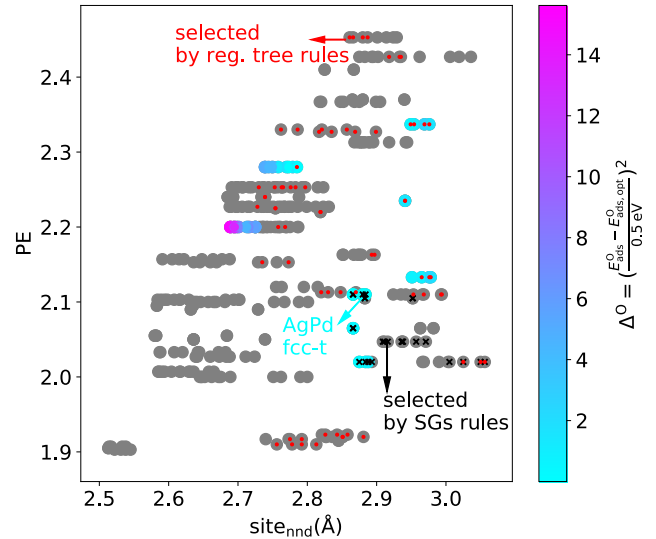

**Figure 3.** SG rules for monometallic surface sites with optimal range of oxygen adsorption energies exploited for the design bimetallic alloys with 1:1 atomic ratio. Representation of the alloy surface sites on the coordinates of two of the identified relevant descriptive parameters. The explicitly calculated data points are colored according to their $\Delta^O$ value. The data points simultaneously selected by the SG rules of Table. S1 (for the target $\Delta^O$) are marked with black crosses. The data points selected by the regression tree are shown as red dots.

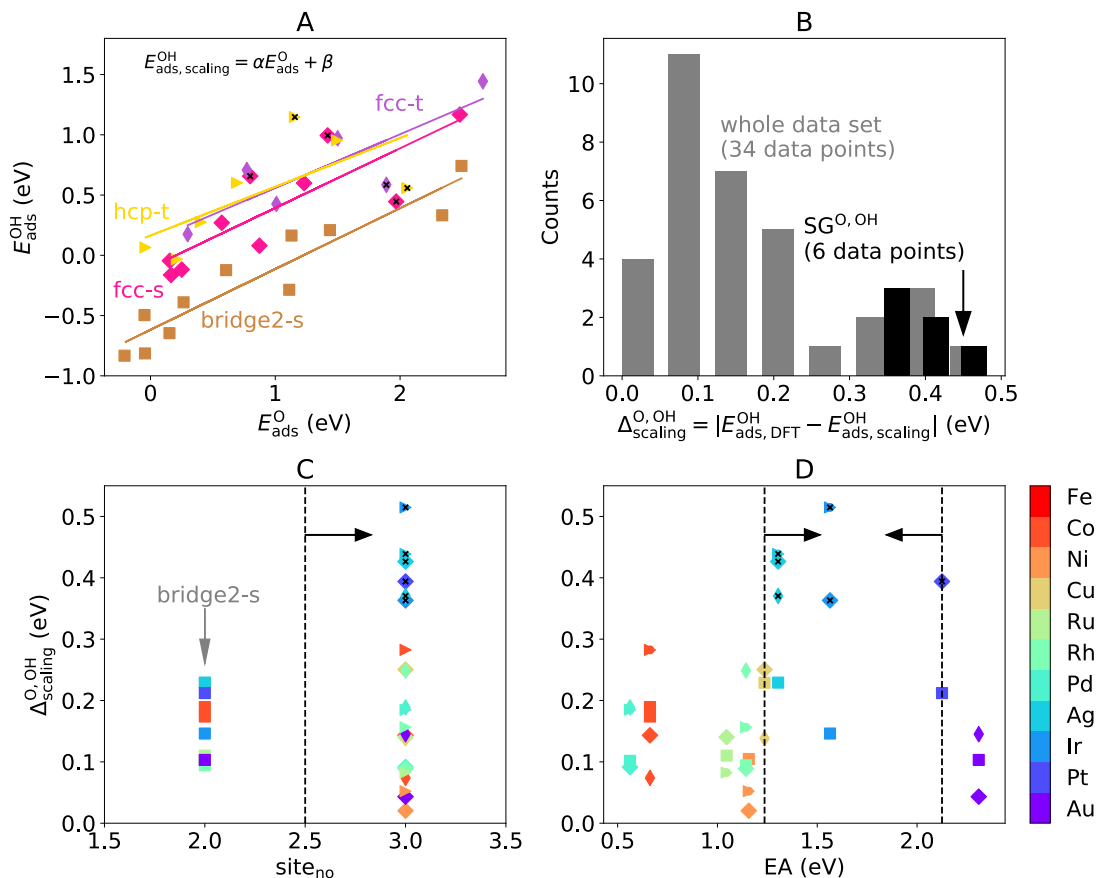

**Figure 4.** SGD of transition-metal catalysts and adsorption sites of fcc (211) surfaces that deviate from the linear scaling relations. A: Scaling relations between oxygen (O) and hydroxyl (OH) species for different adsorption sites of the fcc (211) monometallic surfaces. B: Distribution of the target ($\Delta_{scaling}^{O,OH}$) within the population and in the identified SG. C and D: SG selectors (indicated by the dashed lines and arrows) on the selected key descriptive parameters coordinates: number of atoms in the ensemble ($site_{no}$) and electron affinity (EA), respectively. The data points corresponding to the SG are marked with black crosses in A, C and D.

Figure 3 shows the alloy surface sites (grey circles) on the coordinates of the two descriptive parameters identified by the SG rule shown in (8): $site_{nnd}$ and PE. In this figure, the black crosses indicate the alloy surface sites selected by the SG rules. Surface sites of the following alloy compositions are selected: AgIr, AgPd, AgPt, and RhAg. The $\Delta^O$ values for the surface sites of 4 alloys (37 data points), which were explicitly evaluated by DFT in reference [8], are also shown as colored circles in Fig. 3. The fcc-t site of the AgPd alloy, for which the calculated $\Delta^O$ is the lowest – and equal to zero-, is an outstanding alloy surface site which was correctly selected by the SG rule. These results show how SGD can be used to quickly select, out of many candidate materials, the promising alloys that should be investigated in more detail.

**Subgroups of surface sites deviating from the linear scaling relations between O and OH adsorption energies for the OER**

The linear trends often observed between adsorption energies of different surface species impose, in some reactions, a limit to the maximum performance that can be achieved. This is because the linear scaling relations imply that the absorption of two related species cannot be tuned independently, limiting the possibilities for catalyst optimization. For instance, in the OER, the adsorption energies of the three key intermediates, O, OH, and OOH, are correlated[21] and the O adsorption energy needs be decreased with respect to OOH adsorption energy in order to decrease the limiting potential and thus maximize the performance.[12] To overcome this limitation imposed by the linear scaling relations, an immense effort has been put into strategies to identify exceptional materials and adsorption sites that "break", or deviate from, the scaling relations.[13] This situation calls for local approaches, since most of the materials are typically well-described by the linear approximate model. To illustrate how the SGD can be used to find outstanding surface sites that deviate from linear scaling relationships, we next search for SGs describing fcc (211) surface sites of monometallic surfaces providing high deviations from the scaling relations between atomic oxygen (O) and hydroxyl (OH). For this purpose, we first establish linear models for each adsorption site on which both O and OH present a (meta)stable adsorption: fcc-t, hcp-t, fcc-s and bridge2-s (show in colors in Fig. 4A). These models have the form

$$E_{ads,scaling}^{OH} = \alpha E_{ads,DFT}^{O} + \beta, \quad (9)$$

where $\alpha$ and $\beta$ are fitted coefficients, different for each surface site. In total, 36 data points are used. The linear fits (Fig. 4A) evidence that most of the data points are well described by the scaling relation. Indeed, the deviations from the linear trend are typically lower than 0.2 eV (Fig. 4B). The bridge2-s surface site is in particular well captured by the linear model. We define the quantity

$$\Delta_{\text{scaling}}^{\text{O,OH}} = |E_{\text{ads,DFT}}^{\text{OH}} - E_{\text{ads,scaling}}^{\text{OH}}|, \quad (10)$$

the absolute difference between the OH adsorption energy estimation by the scaling relation ($E_{\text{ads,scaling}}^{\text{OH}}$) and the actual DFT-calculated value ($E_{\text{ads,DFT}}^{\text{OH}}$) as target for the SGD approach. In this way, the interesting data points, i.e., the surface sites that are worst described by the linear trend, correspond to high values of $\Delta_{\text{scaling}}^{\text{O,OH}}$. Most of the observations in the population correspond to low $\Delta_{\text{scaling}}^{\text{O,OH}}$ values (Fig. 4B). We are thus interested in SGs with an overall distribution of the target value as different as possible from the distribution of this quantity in the population. This requirement can be introduced in the SGD by means of the following utility function:

$$u(P, SG) = D_{\text{cJS}}(P, SG). \quad (11)$$

In (11), $D_{\text{cJS}}(P, SG)$ is the cumulative-distribution-function formulation[30] of the Jensen-Shanon divergence between the distribution of the target values in the SG and the distribution of the target values in global population.[30] $D_{\text{cJS}}$ measures the dissimilarity between two distributions. It assumes small values for similar distributions and increases as the distributions have different standard deviations or mean values (see further details in ESI). The candidate descriptive parameters shown in Table 1 are also used here, and only the monometallic systems are initially considered.

The SGD approach identifies a SG containing 6 data points, i.e., ca. 17 % of the population, which is narrow and has relatively high target values with respet to the population (Fig. 4B, in black). Indeed, this SG contains the surface sites deviating the most from the linear scaling relations (Fig. 4A, in which the data points belonging to this SG are shown as black crosses). The sites fcc-s, fcc-t, and hcp-t of the Ag surface, the sites fcc-s, and hcp-t of the Ir surface and the site fcc-s of the Pt surface are in this SG. This SG is defined by the selector

$$\sigma^{\text{O,OH}} = \text{site}_{\text{no}} > 2.5 \land 1.236 \text{ eV} \leq EA \leq 2.125 \text{ eV}, \quad (12)$$

as shown in Fig. 4 C and D. Therefore, the number of atoms in the surface site ensemble ($\text{site}_{\text{no}}$) and the electron affinity of the metal ($EA$) are relevant parameters related to high $\Delta_{\text{scaling}}^{\text{O,OH}}$. The constrain on $\text{site}_{\text{no}}$ excludes the bridge2-s sites from the SG and shows that surface sites composed by more than two atoms, on which the adsorbate can be more highly-coordinated, are more prone to deviate from the linear trend. The conditions on $EA$, in turn, shows that this outstanding behavior is limited to only some of the metals, and this is encoded in this element-dependent (atomic) parameter.

We also exploited the rules defining the SGs of surface sites deviating from the linear scaling relations (see list in Table S1) to address transition-metal alloys (Fig. 5). The comparison of explicitly calculated $\Delta_{\text{scaling}}^{\text{O,OH}}$ on alloys with alloys selected by the SG rules show that the constraints derived based on the monometallic systems correctly indicate which alloys surface sites deviate the most from the scaling relations (purple points). In particular, the alloy surface site with highest $\Delta_{\text{scaling}}^{\text{O,OH}}$ = 0.27 eV, fcc-s AgAu, is part of the identified SG. Thus, the SG rules describing surface sites that deviate from the linear scaling relations are generalizable beyond the data set used for their derivation.

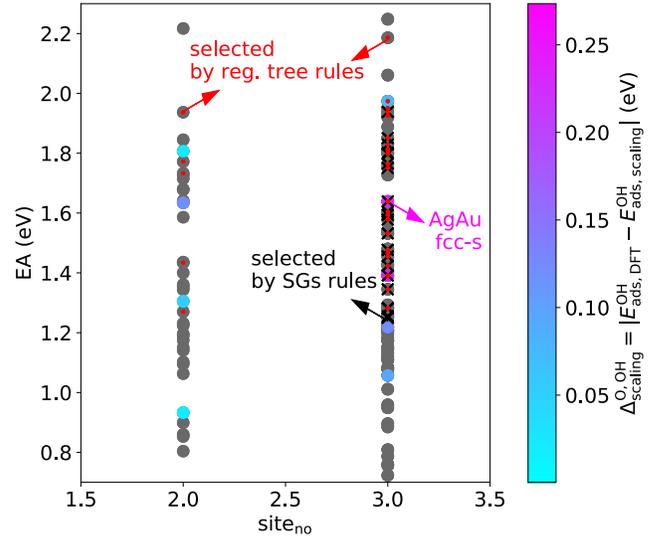

**Figure 5.** SG rules for fcc (211) transition-metal catalysts and adsorption sites deviating from the linear scaling relations exploited for the design bimetallic alloys with 1:1 atomic ratio. Representation of the alloys on the coordinates of the identified relevant descriptive parameters. The explicitly calculated data points are colored according to their $\Delta^{\text{O}}$ value. The data points simultaneously selected by the SG rules of Table. S1 (for the target $\Delta_{\text{scaling}}^{\text{O,OH}}$) are marked with black crosses. The data points selected by the regression tree are shown as red dots.

Overall, our results demonstrate the potential of SGD to detect complex local patterns associated to outstanding behavior. Furthermore, generalizable SG rules were derived based on extremely small data sets compared to those typically needed for widely-used artificial-intelligence methods. This makes the SGD approach useful for several catalysis and materials-science applications in which only small (consistent) data sets are available. This contribution also demonstrates how the sharing of well-annotated FAIR (Findable, Accessible, Interoperable, and Re-purposable) data, increasingly available via common data infrastructures,[31] can enable scientific insights beyond the original purpose for which the data was created and used.

Even though the SGD approach enables the screening of new materials, as demonstrated above, it does not provide predictions of oxygen adsorption energies for each single adsorption site. In particular, the SGD rule does not tell the most stable surface sites for a given surface containing several possible adsorption sites on which oxygen might bind with different strength. However, knowing the relative stability of adsorption configurations might be important for the description of a catalytic process. This is addressed in reference [8] via the symbolic-regression sure-independence-screening-and-sparsifying-operator (SISSO) approach. Indeed, the models derived by SISSO are able to quantitatively describe the adsorption energies for each different material and surface site, thus identifying the most stable adsorption configurations. The SISSO approach allows capturing the potentially nonlinear relationships among the most relevant input parameters that best model the adsorption energies. Contrary to SISSO, however, SGD provides a local description focused only on specific desired behaviors. Furthermore, SGD identifies simple constraints on the most relevant input parameters, which are helpful for rationalizing the possible underlying phenomena. The SGD analysis presented here thus advances the physical understanding of the *local* behavior with respect to the previous SISSO work.

Finally, we note that the dynamic restructuring of the catalyst material that might occur under reaction conditions, thus influencing the surface structure on which the reactions take place,[1,2] is not being taken into account in our analysis. This requires alternative modelling strategies.[32,33,3]

**Comparison of subgroup discovery with decision tree regression**

We also trained regression (reg.) trees[34] using the same data sets of targets and descriptive parameters as for SGD (see details in ESI). Similar to SGD, reg. trees also provide rules describing subsets of data identified during the training. These subsets of data are called "leaves", and reg. trees provide predictions of target values according to the leaf to which a given data point belong, e.g., one prediction per leaf. For the $\Delta^O$ target, the reg. tree approach identified, on the leaf with the minimum predicted value of 0.115 eV, 7 adsorption sites. The rules for this leaf are:

$$\sigma^{O,\text{tree}} = \epsilon_d \leq -1.387 \text{ eV} \wedge \text{site}_{\text{nnd}} \geq 2.651 \text{ Å} \wedge \text{IP} \leq 9.04 \text{ eV} \wedge f_{sp} \geq 1.109 \wedge \text{DOS}_d \leq 1.71 \text{ eV}^{-1} \,. \quad (13)$$

For the $\Delta^{O,OH}_{\text{scaling}}$ target, the reg. tree approach identifies, in the leaf with maximum predicted value of 0.418 eV, 6 adsorption sites. The rules describing this leaf are:

$$\sigma^{O,OH,\text{tree}} = \epsilon_d \leq -1.805 \text{ eV} \wedge \text{PE} \leq 2.41 \wedge \text{EA} \geq 1.27 \text{ eV} \wedge \text{CN} \geq 7.667\,. \quad (14)$$

Interestingly, the same subset of data as the one selected by the SG rules discussed in Fig. 4 is selected by (14).

The reg. tree rules obtained using monometallic surfaces were then applied to select bimetallic alloys. The alloy surface sites selected by the reg. tree rules (13) and (14) are shown as red dots in Fig. 3 and 5, respectively. The reg. tree rule describing low $\Delta^O$ values (Fig. 3) selects several of the alloys systems for which the calculated $\Delta^O$ is relatively low. However, it does not select the fcc-t site of the AgPd alloy, for which the calculated $\Delta^O$ is equal to zero. The reg. tree rule describing high $\Delta^{O,OH}_{\text{scaling}}$ values (Fig. 5) overall selects the surface sites that deviate the most from the scaling relations. However, some of the bridge surface sites of alloys surfaces are also selected, which is in contrast with the relatively low explicitly calculated values of $\Delta^{O,OH}_{\text{scaling}}$ for such sites. In particular, the bridge2-s site of AgAu alloy, selected by the reg. tree rule, has a calculated $\Delta^{O,OH}_{\text{scaling}} = 0.021$ eV (vs. the prediction $\Delta^{O,OH}_{\text{scaling}} = 0.418$ eV). We ascribe the worse performance of the reg. tree approach with respect to SGD for the present data set and targets to the global character of the loss function used to select the rules in reg. trees. Indeed, the loss function minimized during the training is, for reg. tree, the prediction error over the *entire* data set. The few statistically exceptional cases therefore do not significantly impact the choice of rules. In SGD, in contrast, the rule is dictated *solely* by the exceptional data points.

**Conclusions**

In this paper, we applied the SGD approach to identify the most relevant atomic, bulk and surface properties - as well as rules associated to those parameters - describing outstanding SGs of transition-metal surface sites. In particular, we demonstrated this approach using a data set of DFT-calculated adsorption energies[8,24] by searching for surface sites (i) that present optimal range of oxygen binding strength for the ORR or (ii) that deviate the most from linear scaling relations between O and OH adsorption energies that impose a limit to the OER performance. The SGs rules not only hint at the relevant underlying physicochemical processes that govern the local statistically exceptional behavior, but are also suitable for guiding the design of challenging bimetallic alloys.

## Electronic supplementary information

Additional SGD and regression tree details are available as ESI. The SGD analysis described in this publication can be found in a Jupyter notebook at the *NOMAD Artificial-Intelligence Toolkit* (https://nomad-lab.eu/AItutorials/), where it can be repeated and modified directly in a web browser.

## Author information

**Corresponding author**

*foppa@fhi-berlin.mpg.de

**ORCID IDs**

Lucas Foppa: https://orcid.org/0000-0003-3002-062X
Luca M. Ghiringhelli: https://orcid.org/0000-0001-5099-3029

## Acknowledgments

Matthias Scheffler is acknowledged for insightful discussions. We also thank Erwin Lam for critically reading the manuscript.

## Declarations

**Funding**

This project has received funding from the European Union's Horizon 2020 research and innovation program (#951786: The NOMAD European Center of Excellence). L. F. acknowledges the funding from the Swiss National Science Foundation, postdoc mobility grant #P2EZP2_181617 and L. M. G. acknowledges funding from the ERC grant #740233: TEC1p.

**Conflict of interest**

The authors declare no competing interests.

**Availability of data**

All data analyzed in this study are included in this published article as supplementary information files.

**Code availability**

The SGD analysis presented in this paper was performed with CREEDO, a web application that provides an intuitive graphical user interface for real knowledge discovery algorithms and allows to rapidly design, deploy, and conduct user studies. CREEDO is available under http://realkd.org/creedo-webapp/. See also the NOMAD analytics-toolkit for a tutorial.